\documentclass[12pt]{article}
\usepackage[english]{babel}
\usepackage{amsmath,amsthm}
\usepackage{amsfonts}
\usepackage[utf8]{inputenc}
\usepackage{graphicx}

\theoremstyle{definition}

\theoremstyle{remark}

\numberwithin{equation}{section}
\begin{document}

\title{A stability analysis of the static EKG Boson Stars.}%
\author{D.S. Fontanella$^{1}$, A. Cabo$^{2}$\\ \\
\small{$^{1, 2}$\textsl{Theoretical Physics Department,}} \\
\small{\textsl{ Instituto de Cibernética, Matemática y Física,}}\\
\small{\textsl{ Calle E, No. 309, Vedado, La Habana, Cuba.}}\\
\small{$^1$ \texttt{duvier@icimaf.cu} , $^2$ \texttt{cabo@icimaf.cu} }
}%
\maketitle
\begin{abstract}
The stability of the recently proposed static solutions for boson stars is analyzed.
These solutions of Einstein-Klein-Gordon (EKG) equations arise from considering
the interaction of a real scalar field with matter. We assume that the inclusion of the
scalar field in addition with matter, allows to justify that stability implies that the total
mass of the solution should grow when the initial condition for the density of matter
at the origin is also increased. Employing numerical values for the static boson star
based on a linear relation between the source and the energy density and between
this and the pressure, we found the relation that linked the the scalar field at the
origin with the matter energy density (MED) in the same point. We also determine
the behavior of the total mass (TM) with the matter energy density in the origin, by
also obtaining through this and the weak energy condition, two possible ranges for
stable solutions of static boson stars.
\end{abstract}

\section{Introduction}\label{sec1}
The search for stable configurations for compact objects play an important role in contemporary Astrophysics. There is a great variety of theoretically proposed objects of the most diverse natures \cite{MM2004}, \cite{jetzer1992},~ \cite{2008PhR...467..117S}, \cite{2005ICRC....3..125M} which continue being studied. Boson stars are ones of the most widely accepted configurations assumed for exploring the possibility of the existence of such compact objects. In the article \cite{cabo2020} a static boson star configuration was found solving the EKG equations for a real scalar field, with spherical symmetry  being coupled with matter. These solutions were found numerically by introducing a source being proportional to the energy density of the matter and looking for the value of the scalar field at the origin that would determine a Yukawa potential at large distance, for the scalar field. A main result of that work is to show that the inclusion of the interaction of the scalar field with matter allows for the existence of static solutions of the EKG equations, a property that these equations lack  \cite{jetzer1992}.\\
The present work is composed of the following parts, first we present the system of equations that were solved in our previous paper and that determine the boson star solution. In second place, an investigation of the stability criterion is carried out. The relation between matter energy density at the origin and the scalar field also in the origin,  which determine the existence of the solutions, is evaluated . Finally, we find the TM vs MED in the origin plot which combined with the weak energy condition, results in two possible ranges for stable solutions of the configuration.

\section{The field equations}
Consider a metric defined by the following squared interval and
coordinates%
\begin{flalign}
&ds^{2}=\mathit{v}(\rho){dx^{o}}^{2}-u(\rho)^{-1}d\rho^{2}-\rho
^{2}(sin^{2}\theta\text{ }d\varphi^{2}+d\theta^{2}),\\
&x^{0}=c\text{ }t\text{, \ \ \ }x^{1}=\rho,\\
&x^{2}=\varphi,\text{ \ }x^{3}\equiv\theta,
\end{flalign}
and an energy-momentum tensor for a real scalar field in interaction with matter:%
\begin{align}
T_{\mu}^{\nu}  &  =-\frac{\delta_{\mu}^{\nu}}{2}(g^{\alpha\beta}{\Phi
}_{,\alpha}{\Phi}_{,\beta}+m^{2}{\Phi}^{2}+2\text{ }J(\rho)\text{ }%
\Phi)\nonumber\\
&  \text{ \ \ \ }+g^{\alpha\nu}{\Phi}_{,\alpha}{\Phi}_{,\mu}+P\text{ }%
\delta_{\mu}^{\nu}+u^{\nu}u_{\mu}(P+e).
\end{align}
After making several coordinate changes as detailed in \cite{cabo2020}, the EKG system can be reached:

\begin{subequations}
\begin{align}
\begin{split}
&{\frac{u^{\prime}(r)}{r}}-{\frac{1-u(r)}{r^{2}}}  =-\frac{1}{2}%
(u(r)\phi^{\prime}{(r)}^{2}+\phi(r)^{2} \\ 
&+2j(r)\phi(r))-e(r),\label{ec1}\\
\end{split}\\
\begin{split}
&\frac{u(r)}{v(r)}\frac{v^{\prime}(r)}{r}-{\frac{1-u(r)}{r^{2}}} =-\frac
{1}{2}(-u(r){\phi}^{\prime}{(r)}^{2}+\phi(r)^{2}\\ 
&+2j(r)\phi(r))+p(r),\label{ec2}\\
\end{split}\\
\begin{split}
&p^{\prime}(r)+(e(r)+p(r))\frac{v^{\prime}(r)}{2v(r)}-\phi(r)\text{
}j^{\prime}(r) =0,\label{ec3}\\
\end{split}\\
\begin{split}
&j(r)+{\phi}(r)-u(r)\text{ }{\phi}^{\prime\prime}(r)={\phi}^{\prime
}(r){\Large (}\frac{u(r)+1}{r}-r\text{ }(\frac{{\phi}(r)^{2}}{2}\\ 
&+j(r){\phi}(r)+\frac{e(r)-p(r)}{2}){\Large )},\label{ec4}
\end{split}
\end{align}
\label{straincomponent}
\end{subequations}
\index{non}
where  $\phi(r)$ is the scalar field, $j(r)$ is the source and  $e(r)$, $p(r)$ the energy density and the matter pressure respectively. Note that we used the Bianchi identity too for completing the system of the equations. So, in general the set of equations is composed as fallow: The two first equations are the Einstein one referent to the two first component of the metric $v(r)$ and $ u(r)$. The other one, \eqref{ec3} is the dynamic equation for the energy, the pressure and the scalar field, and it substitutes the two Einstein
equations that are associated to both angular directions. This relation results from the Bianchi identity. Finally, \eqref{ec4} is the Klein-Gordon equation for the scalar fields.

This system describes the dynamics of the space time for a configuration of scalar field and matter coupled through the source $j(r)$ .\\ Let us fix the boundary conditions at a very small value $\delta = 10^{-6}$ around the origin in the following way:

\begin{align}
u(\delta)  &  =1,\\
v(\delta)  &  =1,\\
\phi(\delta)  &  =\phi_{0},\\
e(\delta)  & = e_{0},
\end{align}
and assume a linear relationship between the source and the matter energy density as well as between the pressure and the same matter energy density as
 $$j(r) = g ~e(r),$$  $$p(r) = n ~e(r).$$
In the work \cite{cabo2020} the solutions were determined by setting a value of the energy density of the matter at the origin and determining the value of the scalar field also at the origin in such a way that this field takes the form of a Yukawa potential at large radius. In this way a whole family of regular solutions can be found.

\subsection{Stability criterion}
Now we will impose the restrictions placed by the weak energy condition and the fact that the total mass of the system must increase when the value of the energy density is also increased at the origin \cite{shapiro}. These  criteria will be examined here for the entire family of solutions, determining those that meet them. These conditions will be easily checked if we start from considering the formula for the total mass

\begin{align}
M(e_{0})  &  =\int dr\text{ }4\pi\text{ }r^{2}\text{ }e
_{t}(r),\\
e_{t}(r)  &  =\frac{1}{2}(u(r)\phi^{\prime}{(r)}^{2}+\phi
(r)^{2}+2j(r)\phi(r))+e(r).\nonumber
\end{align}

\begin{figure}[h!]
	\centerline{\includegraphics[width=.8\linewidth]{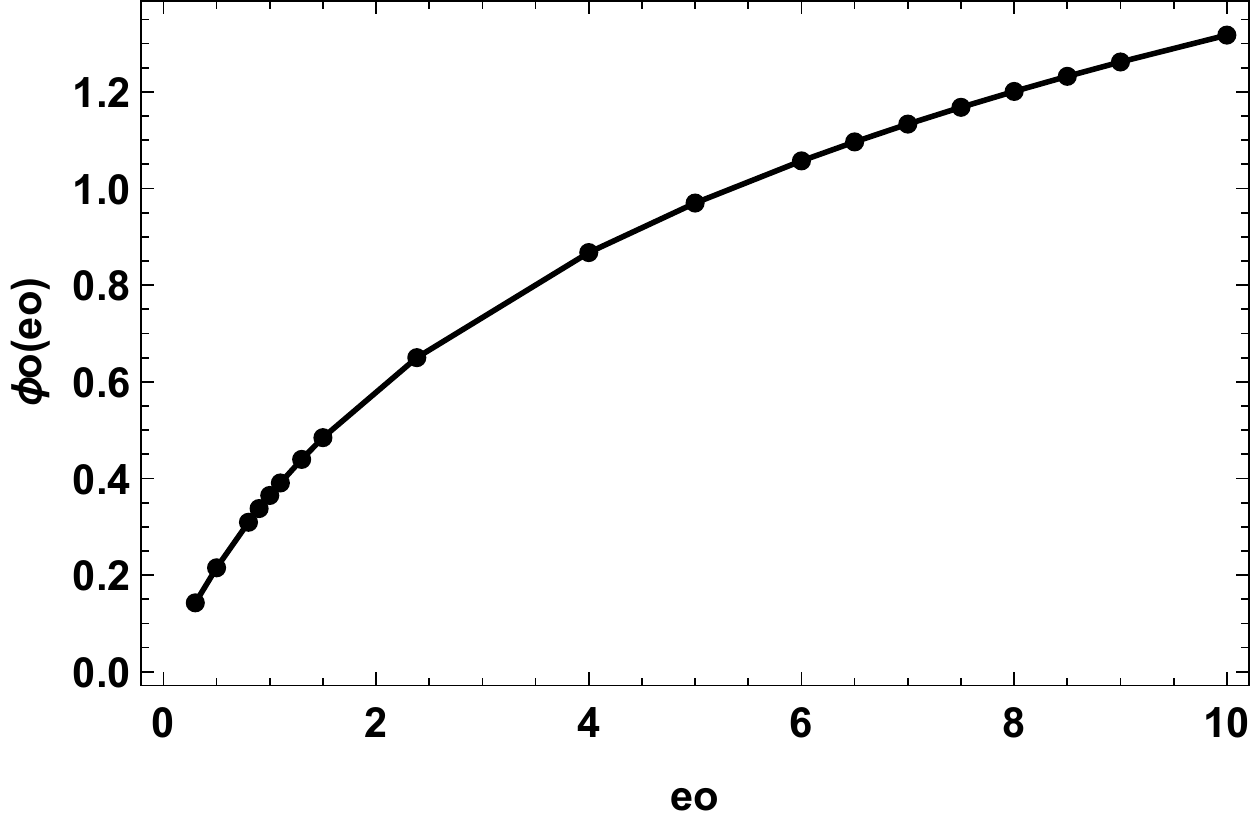}}
	\caption{Relation between matter energy density and the scalar field at the origin~($\phi_{0}$) needful for obtaining  plausible solutions. The relation was found fixing $g = 0.9$, ~$n = 0.025$. \label{campoSvseo}}
\end{figure}

As stated in \cite{cabo2020} the value of the initial field at the origin is a function of the matter energy density at the origin. In other words, depending on the density of matter that the object in question has, it will need a specific value of scalar field at the origin in order to show a Yukawa like behaviour of the scalar field in the faraway regions. This relationship was determined using Mathematica software version 11.1 and the results are shown in the table \ref{valores}. The function $\phi_{0}(e_ {0})$ defined by the table is shown in the figure \ref{campoSvseo}.

\begin{center}
\begin{table}[h!]%
\centering
\caption{The scalar field and total mass values corresponding to the fixed matter energy density at the origin.\label{valores}}%
\begin{tabular*}{20pc}{@{\extracolsep\fill}lcc@{\extracolsep\fill}}%
\textbf{$e_{0}$} & \textbf{$\phi_{0}$} & \textbf{$M|_{r=10}$} \\
0.3 & 0.142664736127 & 16.7838 \\
0.5 & 0.215184567562 & 16.9132 \\
0.8 & 0.309367889200 & 17.149\\
0.9 & 0.337735444700 & 17.1977 \\
1.0 & 0.364823000000 & 17.2341\\
1.1 & 0.390743919420 & 17.2535 \\
1.3 & 0.439437983840 & 17.2678 \\
1.5 & 0.484445641130 & 17.2546 \\
2.0 & 0.583890263504 & 17.1616 \\
4.0 & 0.867447339143 & 16.7466 \\
5.0 & 0.969939323350 & 16.6329\\
6.0 & 1.057176196658 & 16.5746\\
6.5 & 1.096426605833 & 16.5623 \\
7.0 & 1.133275616740 & 16.5595 \\
7.5 & 1.168018636795 & 16.5651\\
8.0 & 1.200900067643 & 16.578\\
8.1 & 1.207271535456 & 16.5814 \\
8.15& 1.210432922380 & 16.5832 \\
8.2 & 1.21357829962  & 16.5851 \\
8.3 & 1.219821748053 & 16.589 \\
8.5 & 1.232124212860 & 16.5976 \\
9.0 & 1.261863522592 & 16.6232\\
10.0& 1.317454277700 & 16.6899\\
\end{tabular*}
\end{table}
\end{center}
Note how that, if the MED at the origin increases the scalar field needful to obtain a plausible solution increases too. Doing a fit for the set of points in  (table\ref{valores}[1-2]) we obtain that the functional relation between scalar field and the MED can be expressed as
$$\phi_{0}[e_{0}]=1.86882\sqrt[5]{e_{0} + 0.536945}-1.66797.$$

\section{Stability Regions}

Firstly, after solving the \eqref{ec1}-\eqref{ec4} system of equations  and fixing  the values of the MED at the origin we find the value of the scalar field at the origin, such that the field behaves like a Yukawa potential for large distances.
\begin{figure}[h!]
	\centerline{\includegraphics[width=.8\linewidth]{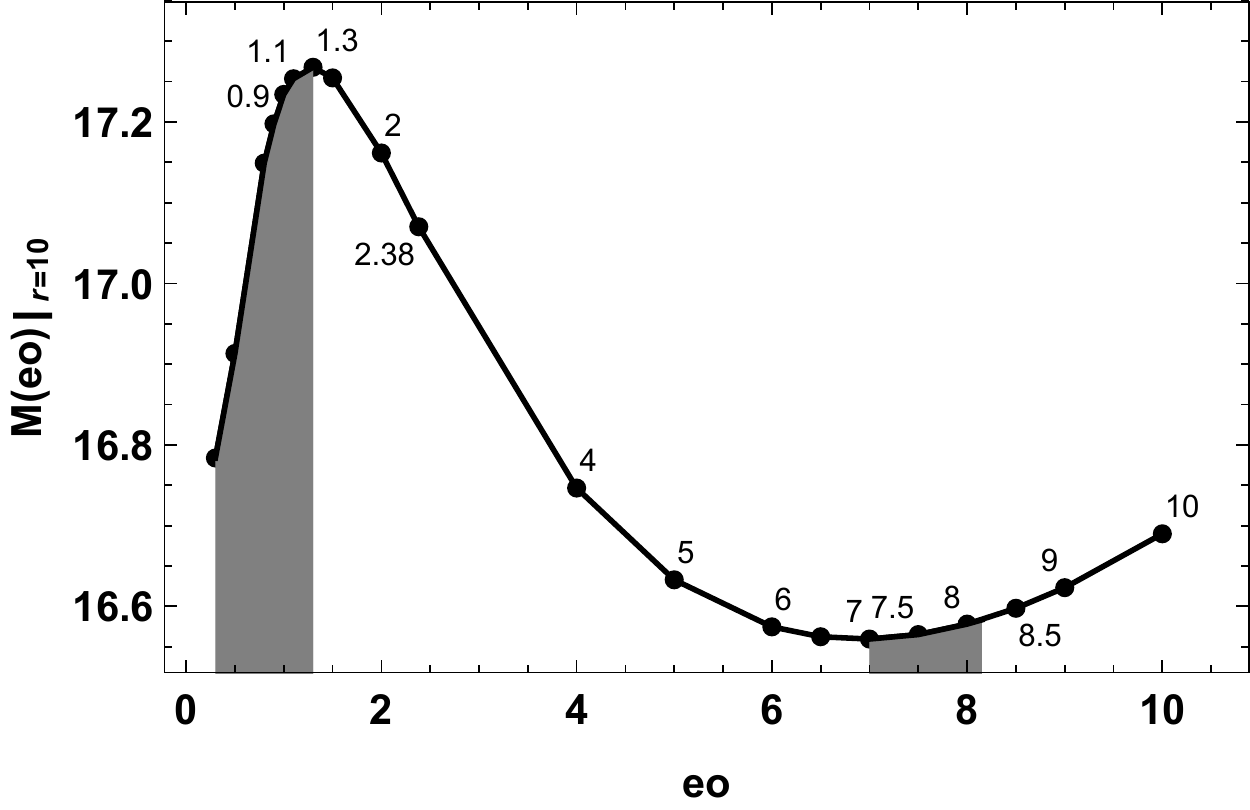}}
	\caption{ Plot of the total mass vs matter energy density at the origin for the values $g = 0.9$ ,~$n = .025$. The dark zones are the stable ones. Note how in spite of having a positive slope for $e_0$ values over $8.15$ the system is unstable  because the total energy density starts to have a negative region.\label{MvsEo}}
\end{figure}
Afterwards, by integrating the total energy from $r = \delta$ to $r = 10$, at which the total energy density is approximately close to zero, we find the total mass vs MED curve as shown in figure \ref{MvsEo}. \\

\begin{figure}[h!]
	\centerline{\includegraphics[width=.8\linewidth]{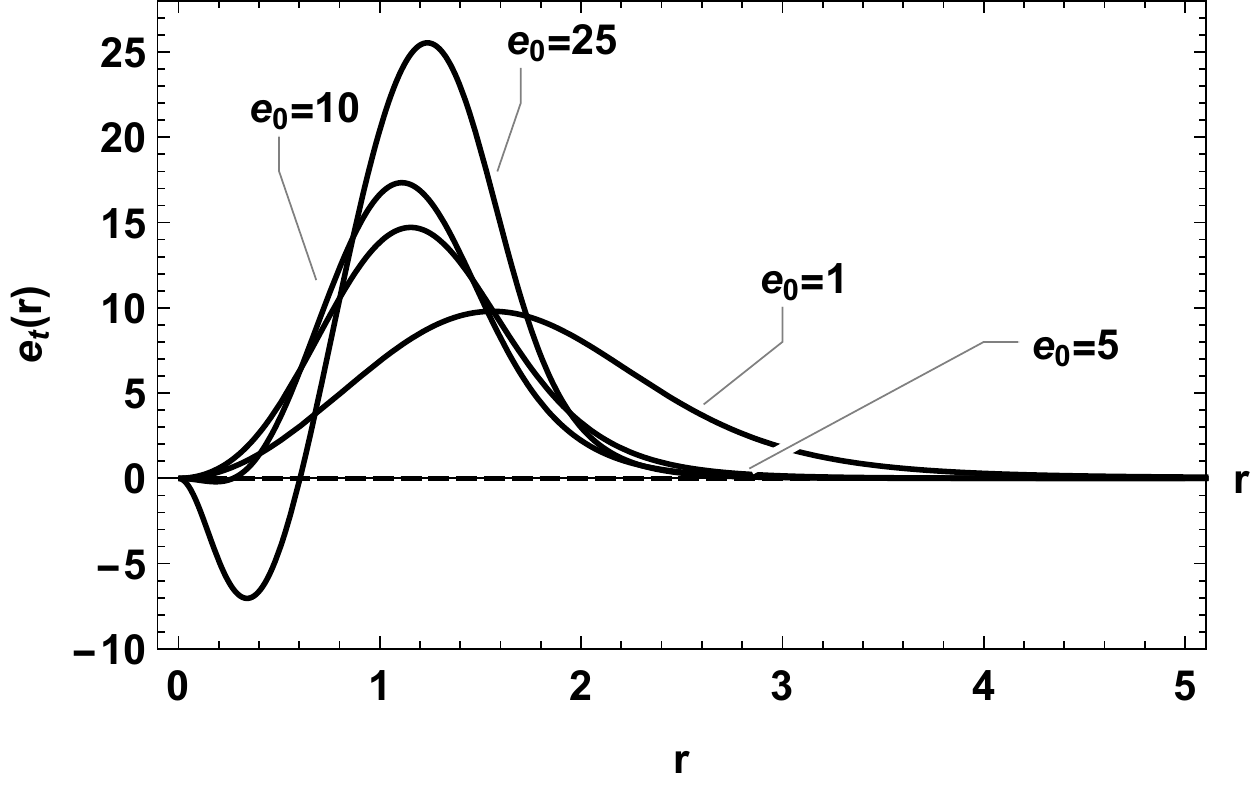}}
	\caption{Behavior of the total energy density with the radial distance from the origin for different values of $e_{0}$ .\label{ETvsRT}}
\end{figure}
\begin{figure}[h!]
	\centerline{\includegraphics[width=.8\linewidth]{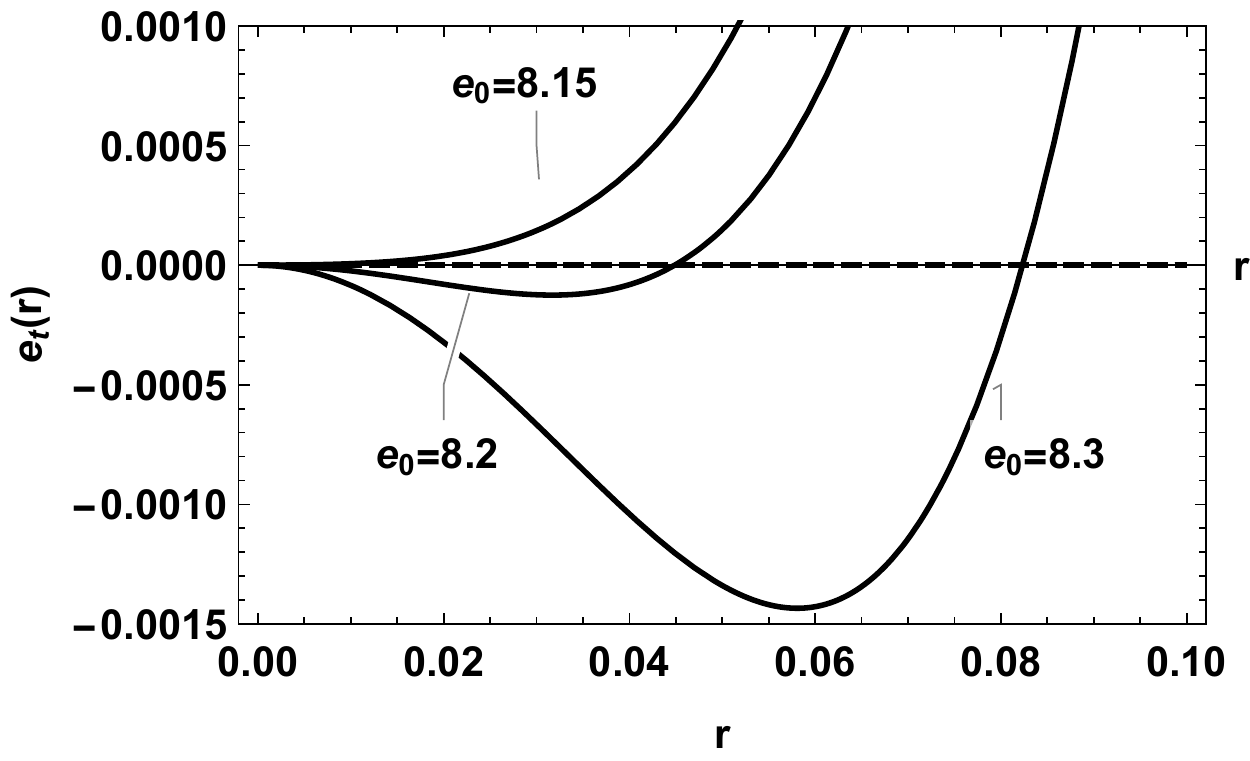}}
	\caption{Behavior of the total energy density with the radial distance from the origin around the value $e_{0}=8.16$.\label{ETvsR}}
\end{figure}
Note that the total mass of the system initially increases with the corresponding increase in the MED at the origin, until it reaches a maximum of $ 17.2678$ at the point $ 1.3 $. From this point it begins to decrease until it reaches a minimum of $ 16.5595 $ in $ e_{0} = 7, $ where it starts to increase again, this time with a lower slope. Then, the slope remains positive for the intervals $ e_{0} <1.3 $ and $ e_{0}> 7 $. Let us analyze the energetic conditions in such ranges. Calculating the total energy density $e_{t}$ for few values of $e_0$, we obtained the plots shown in the figure \ref{ETvsRT}. As it can be observed, for certain values of the energy density of matter at the origin, the total energy begins to have negative values \footnote{For a better illustration, the graphs were only taken for certain values of $ e_{0} $.}. The value of the total energy density at which the curves begin to have a negative region was determined graphically to be $ e_{0} = 8.15 $ in figure \ref{ETvsR}. Also, there  are regions where, despite the fact that the derivative of the total mass with respect to the energy is positive,  the physical system  is not meeting the weak energy condition. Hence, for the system \eqref{ec1}-\eqref{ec4} the stability regions $ e_ {0} <1.3 $ and $ 7 <e_ {0} <8.15 $ result, which are shown as the shaded areas in figure \ref{MvsEo}.

\section{Conclusions}\label{sec5}

We study the stability of the static solutions of the EKG equations for a real scalar field with spherical symmetry and including matter. This analysis was carried out under two simple criteria: The weak energy condition and the need for growth of the mass with the increase in the energy density at the origin \cite{shapiro}. Under these criteria and for a specific configuration described in detail in \cite{cabo2020}, two stability zones were determined, as well as the relationship between the scalar field and the matter energy density at the origin. The analysis will be extended for other values of the pressure and energy relationships as well as for the coupling between the scalar field and matter. In particular, the plans also include studying polytropic relationships.

\section*{Acknowledgments}
The authors acknowledge the partial support of the Office of External Activities of ICTP (OEA), through
the Network on Quantum Mechanics, Particles and Fields (Net-09).


\end{document}